\documentclass[10pt]{article}

\usepackage[utf8]{inputenc}
\usepackage{todonotes}

\usepackage[margin=1in]{geometry}
\usepackage[affil-it]{authblk}
\usepackage{titlesec}
\titlespacing*{\section}{0pt}{1em}{1em}
\usepackage{amsmath}
\usepackage[]{siunitx}
\usepackage[version=4]{mhchem}

\usepackage{natbib}
\setcitestyle{numbers, square}

\title{Exceeding the Shockley-Queisser limit within the detailed balance framework}

\author{Marnik Bercx$^*$, Rolando Saniz, Bart Partoens and Dirk Lamoen}

\affil{EMAT \& CMT groups, Department of Physics, University of Antwerp, Groenenborgerlaan 171, 2020 Antwerp, Belgium}

\date{}

\begin{document}

\maketitle
\begin{abstract}
The Shockley-Queisser limit is one of the most fundamental results in the field of photovoltaics. Based on the principle of detailed balance, it defines an upper limit for a single junction solar cell that uses an absorber material with a specific band gap. Although methods exist that allow a solar cell to exceed the Shockley-Queisser limit, here we show that it is possible to exceed the Shockley-Queisser limit without considering any of these additions. Merely by introducing an absorptivity that does not assume that every photon with an energy above the band gap is absorbed, efficiencies above the Shockley-Queisser limit are obtained. This is related to the fact that assuming optimal absorption properties also maximizes the recombination current within the detailed balance approach. We conclude that considering a finite thickness for the absorber layer allows the efficiency to exceed the Shockley-Queisser limit, and that this is more likely to occur for materials with small band gaps.
\end{abstract}
\vspace{1em}
  
{
  \renewcommand{\thefootnote}%
    {\fnsymbol{footnote}}
  \footnotetext[1]{E-mail: marnik.bercx@uantwerpen.be}
}

\section{Introduction}\label{sec:intro}

Materials play a central role in the effort to produce cheaper and more efficient solar cells. The discovery of improved absorber materials has the potential to significantly increase the cost-effectiveness of photovoltaic devices, but experimental trial and error methods are often slow and expensive. Here, computational material modeling can provide a valuable assist to the material design process, by screening groups of materials for those that have the best properties. 

The Shockley-Queisser limit~\cite{Shockley1961DetailedCells} is one of the most well-known metrics to determine the maximum efficiency an absorber material can produce in a single-junction solar cell. It was proposed in 1961 and provides a direct relation between the band gap of a material and its maximum possible efficiency. More recently, Yu and Zunger expanded on the work of Shockley and Queisser by introducing the Spectroscopic Limited Maximum Efficiency~\cite{Yu2012IdentificationMaterials} (SLME), which takes the absorption coefficient and thickness into consideration for the calculation of the maximum efficiency. The SLME has since been used to investigate the potential of photovoltaic absorber materials such as perovskites~\cite{Meng2016AlloyingApplication}, direct band gap silicon crystals~\cite{Lee2014ComputationalCrystals}, chalcogenides, and other materials. In our recent work on CuAu-like~\cite{Bercx2016First-principlesSilicon} and Stannite~\cite{Sarmadian2016First-principlesChalcogenides} structures, we also used the SLME to study the efficiency of these materials in the context of thin film solar cells. Interestingly, we found several materials with an SLME above the Shockley-Queisser limit, and identified that this is due to the lower recombination current obtained for the material at lower thicknesses.

Since its conception, numerous methods have been proposed to exceed the Shockley-Queisser limiting efficiency~\cite{Nelson2013ExceedingConversion}. Examples include multi-junction~\cite{Shah2004Thin-filmTechnology,Heremans2009StrategiesArchitecture} and hot carrier solar cells~\cite{Konig2010HotDesign}, as well as concepts that use multiple exciton generation~\cite{Hanna2006SolarAbsorbers}. None of these concepts, however, are implemented in the SLME. In this paper, we use a model approach to demonstrate that it is possible to exceed the Shockley-Queisser limit within the detailed balance framework. Simply by dropping the assumption of an infinite absorber layer, i.e. by replacing the Heaviside step function for the absorptivity by a sigmoid function, we obtain efficiencies above the Shockley-Queisser limit. Finally, we analyze for which band gap range a material's efficiency is more likely to exceed the Shockley-Queisser limit.

\section{Shockley-Queisser limit}\label{sec:SQ}

The maximum efficiency $\eta$ is defined as the maximum output power density $P_m$ divided by the total incoming power density from the solar spectrum $P_{in}$:
\begin{equation}
\eta = \frac{P_m}{P_{in}}
\end{equation}
To calculate $P_m$, the power density $P = JV$ is maximized versus the voltage $V$, where the current density\footnote{Note that these current densities are not defined in the conventional way. Rather, they are considered as currents per surface area of the solar cell. This allows us to ignore the surface area of the solar cell in our discussion.} $J$ is derived from the ideal $J-V$ characteristic of an illuminated solar cell:
\begin{equation}
J = J_{sc} - J_0 \left(e^{\frac{eV}{k_B T}} - 1\right),
\end{equation}
where $k_B$ is Boltzmann's constant, $e$ is the elementary charge and $T$ is the temperature of the solar cell. The short-circuit current density $J_{sc}$, also known as the photogenerated current or the illuminated current, is calculated from the number of photons of the solar spectrum that are absorbed by the solar cell:
\begin{equation}
J_{sc} = e \int_0^{\infty} a(E) \Phi_s (E) dE, \label{eq:Ish}
\end{equation}
where $a(E)$ is the absorptivity and $\Phi_s(E)$ is the photon flux density of the solar spectrum. In their original paper, Shockley and Queisser used a blackbody spectrum of $T_s = 6000~\si{\kelvin}$, but the current convention is to use the AM1.5G solar spectrum~\cite{2012ASTMPA}. 

The reverse saturation current density $J_0$ is calculated by considering the principle of detailed balance, i.e. in equilibrium conditions the rate of photon emission from radiative recombination must be equal to the photon absorption from the surrounding medium. Because the cell is assumed to be attached to an ideal heat sink, the ambient temperature is assumed to be the same as that of the solar cell. Hence, the spectrum of the surrounding medium is that of a black body at cell temperature $T$:
\begin{align}
J_0 &= e \pi \int_0^{\infty} a(E) \Phi_{bb}(E) dE\nonumber \\
&= e \pi \int_0^{\infty} a(E) \frac{2E^2}{h^3 c^2} \frac{dE}{e^{\frac{E}{k_B T}}-1}, \label{eq:I0}
\end{align}
where $h$ is Planck's constant and $c$ is the speed of light. Because of its connection with the recombination of electron-hole pairs at equilibrium, $J_0$ is also referred to as the recombination current density~\cite{Cuevas2014TheJ0}. This is the convention we will use here.

To obtain the Shockley-Queisser or detailed balance \textit{limit}, Shockley and Queisser made the assumption that the probability of a photon with an energy above the band gap being absorbed by the cell is equal to unity. This corresponds mathematically to setting $a(E)$ to the Heaviside step function, or, from a physical perspective, to considering an infinitely thick absorber layer. Note that in the original expressions, Shockley and Queisser also included a geometrical factor. However, because we assume the solar cell to have a perfect antireflective coating, as well as a reflective back surface, the geometrical factor is equal to unity~\cite{Ruhle2016TabulatedCells}.

\section{Spectroscopic Limited Maximum Efficiency}\label{sec:SLME}

Shockley and Queisser's detailed balance limit is considered to be one of the most important results in photovoltaic research. However, as a metric for thin film solar cells, it is somewhat limited in its effectiveness, because it only depends on the band gap of the absorber material in the solar cell. In an attempt to find a more practical screening metric, Yu and Zunger introduced the Spectroscopic Limited Maximum Efficiency~\cite{Yu2012IdentificationMaterials} (SLME) in 2012. The SLME differs from the detailed balance limit in two ways. First, the absorptivity $a(E)$, taken as a Heaviside step function in the calculation of Shockley and Queisser, is replaced by the absorptivity $a(E) = 1 - e^{-2\alpha(E)L}$, where $L$ is the thickness and $\alpha(E)$ is the absorption coefficient, calculated from first principles. This allows us to use the SLME to study the thickness dependence of the efficiency, an important tool in the study of thin film solar cells.

Second, the SLME also considers the non-radiative recombination in the solar cell by modeling the fraction\footnote{Actually, Shockley and Queisser also considered the fraction of radiative recombination in their approach. They did not, however, provide a model to calculate it, simply observing that the maximum efficiency is significantly reduced for small fractions $f_r$.} of radiative recombination as a Boltzmann factor, i.e. $f_r = e^{-\frac{\Delta}{kT}}$, with $\Delta = E_g^{da} - E_g$, where $E_g$ and $E_g^{da}$ are the fundamental and direct allowed band gap, respectively. The total recombination current density is then calculated by dividing the radiative recombination current density (Eq.~\ref{eq:I0}) by the fraction of radiative recombination. In this work, we only study direct band gap materials (i.e. $E_g = E_g^{da}$), and hence only radiative recombination is considered ($f_r = 1$), just as in the standard calculation of the detailed balance limit.

\begin{figure}[ht]
\centering
\includegraphics[width=0.8\textwidth]{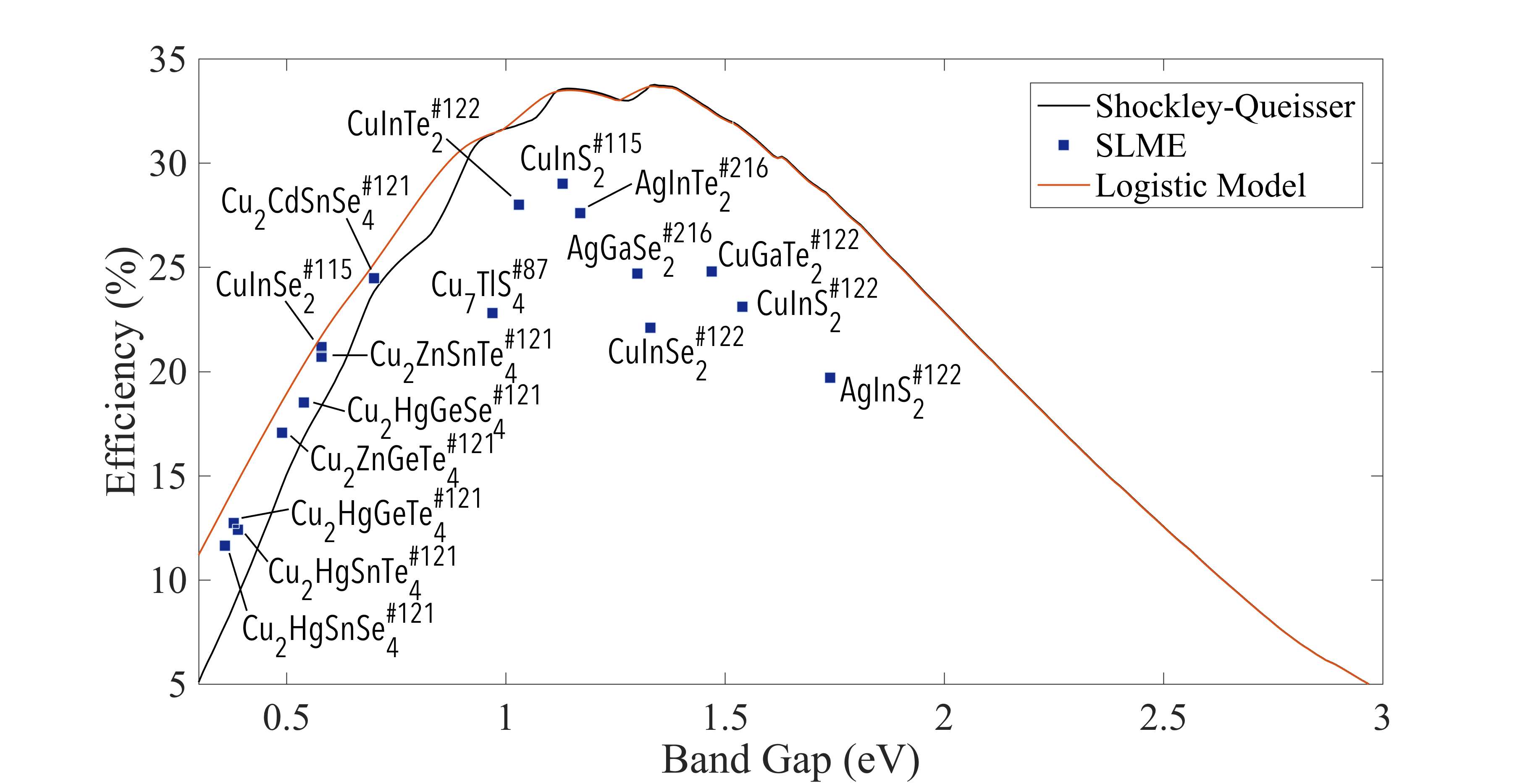}
\caption{\label{fig:SLME}Collection of calculated SLME values from Yu and Zunger~\cite{Yu2012IdentificationMaterials}, as well as our previous work on CuAu-like~\cite{Bercx2016First-principlesSilicon} and Stannite~\cite{Sarmadian2016First-principlesChalcogenides} structures. We have added the space group of the material structure as a superscript. The efficiency values were calculated for a thickness of 0.5~\si{\micro\meter}. The orange curve represents the maximum efficiencies obtained using the logistic model explained in Section~\ref{sec:logistic}.}
\end{figure}

The SLME has been used to investigate the potential of several classes of photovoltaic absorber materials. In Fig.~\ref{fig:SLME}, we show a selection of calculated efficiencies of direct band gap materials from previous work~\cite{Yu2012IdentificationMaterials,Bercx2016First-principlesSilicon,Sarmadian2016First-principlesChalcogenides}, compared with the Shockley-Queisser limit.  We can see that materials typically used in thin-film photovoltaic cells, e.g. chalcopyrite phase \ce{CuIn(S,Se)2}, have a high calculated efficiency. We also note other materials that are less studied with high efficiencies, such as CuAu-like phase \ce{CuInS2} and chalcopyrite phase \ce{CuInTe2}. Most importantly, however, we can see that a significant amount of the presented materials have a calculated efficiency above the Shockley-Queisser limit. Since the calculation of the SLME does not introduce any of the concepts that would typically allow its value to exceed the Shockley-Queisser limit, these results show that for thin-film materials the Shockley-Queisser limit does not necessarily represent an upper limit for the efficiency.

In fact, Shockley and Queisser considered their metric as the detailed balance \textit{limit} because of the assumption that since the step function represents the highest possible absorption spectrum for a material with a specific direct band gap, the resulting efficiency must represent an upper limit. However, as we demonstrated in our previous work~\cite{Bercx2016First-principlesSilicon}, this also means the the recombination current density $J_0$ (Eq.~\ref{eq:I0}) will be maximal. Since electron-hole recombination results in a loss of electrons contributing to the external current, this has a negative effect on the photovoltaic conversion efficiency. Hence, it is possible that there is an absorptivity function that would result in a higher efficiency than the Shockley-Queisser limit. As we can see in Fig.~\ref{fig:SLME}, this is exactly what happens for the presented smaller band gap materials.

\section{Logistic Function Model}\label{sec:logistic}

\begin{figure}[h!]
\centering
\includegraphics[width=0.8\textwidth]{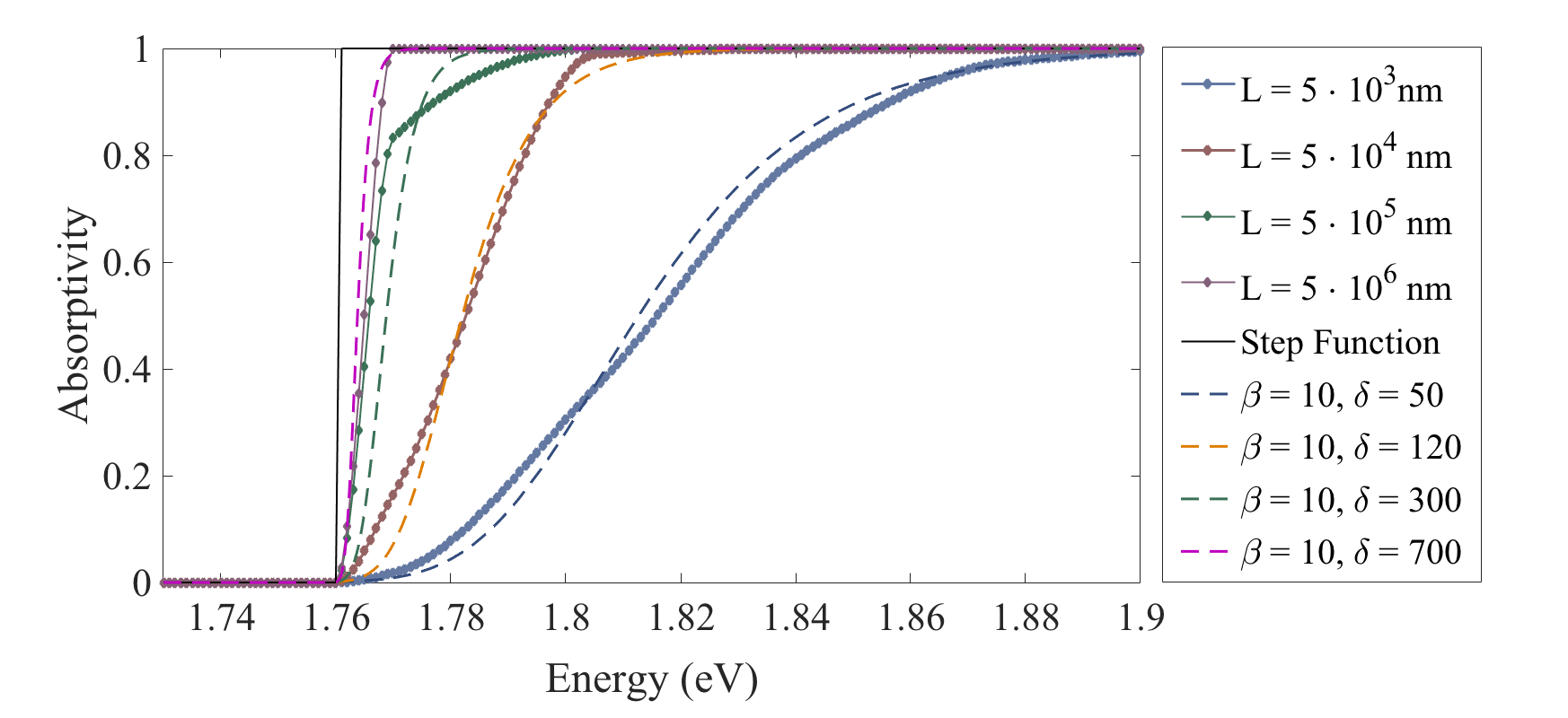}
\caption{\label{fig:step}Comparison of the model function with calculated absorptivity spectra for \ce{Cu2ZnGeS4} at different thicknesses $L$. We can see that the model function shape matches that of the calculated absorptivity quite well as $L,\delta \rightarrow \infty$.}
\end{figure}

The next questions are how far we can exceed the Shockley-Queisser limit, and at which band gaps a material is more likely to do so. Clearly, this will depend on the shape of the absorptivity function. In Fig.~\ref{fig:step}, we show the calculated absorptivity of \ce{Cu2ZnGeS4} for various thicknesses, derived from the absorption coefficient calculated from first principles (For computational details, we refer the reader to~\cite{Sarmadian2016First-principlesChalcogenides}). We can see that the absorptivity has a shape reminiscent of a sigmoid function. In order to analyze the maximum efficiency for materials with a direct band gap in the range 0.3-3~\si{\electronvolt}, we model $a(E)$ using a generalized logistic function:
\begin{equation}
a(E) = f(E) = \frac{1}{(1+e^{-\delta (E - E_g)})^{\beta}},
\end{equation}
where $E_g$ is the band gap of the material, and $\beta$, $\delta$ are parameters that determine the shape of the function. In this model for the absorptivity, the parameter $\delta$ is related to the thickness of the material, as for $\delta\rightarrow\infty$, $f(E)$ approaches the Heaviside step function (Fig.~\ref{fig:step}). The second parameter ($\beta$) is important to make sure that the model function ``starts'' at the band gap, i.e. that its value for $E < E_g$ is suitably small, so that it can be approximated to zero. Since $f(E_g) = \frac{1}{2^\beta}$, and $f(E) < f(E_g)$ for $E < E_g$, increasing $\beta$ to a suitably large value gives us this desired function trait. Here, we choose $\beta = 10$ and set $f(E) = 0$ for $E \leq E_g$. As is clear from Fig.~\ref{fig:step}, this model function describes the shape of the calculated absorptivity spectra quite well.

\begin{figure}[h!]
\centering
\includegraphics[width=0.8\textwidth]{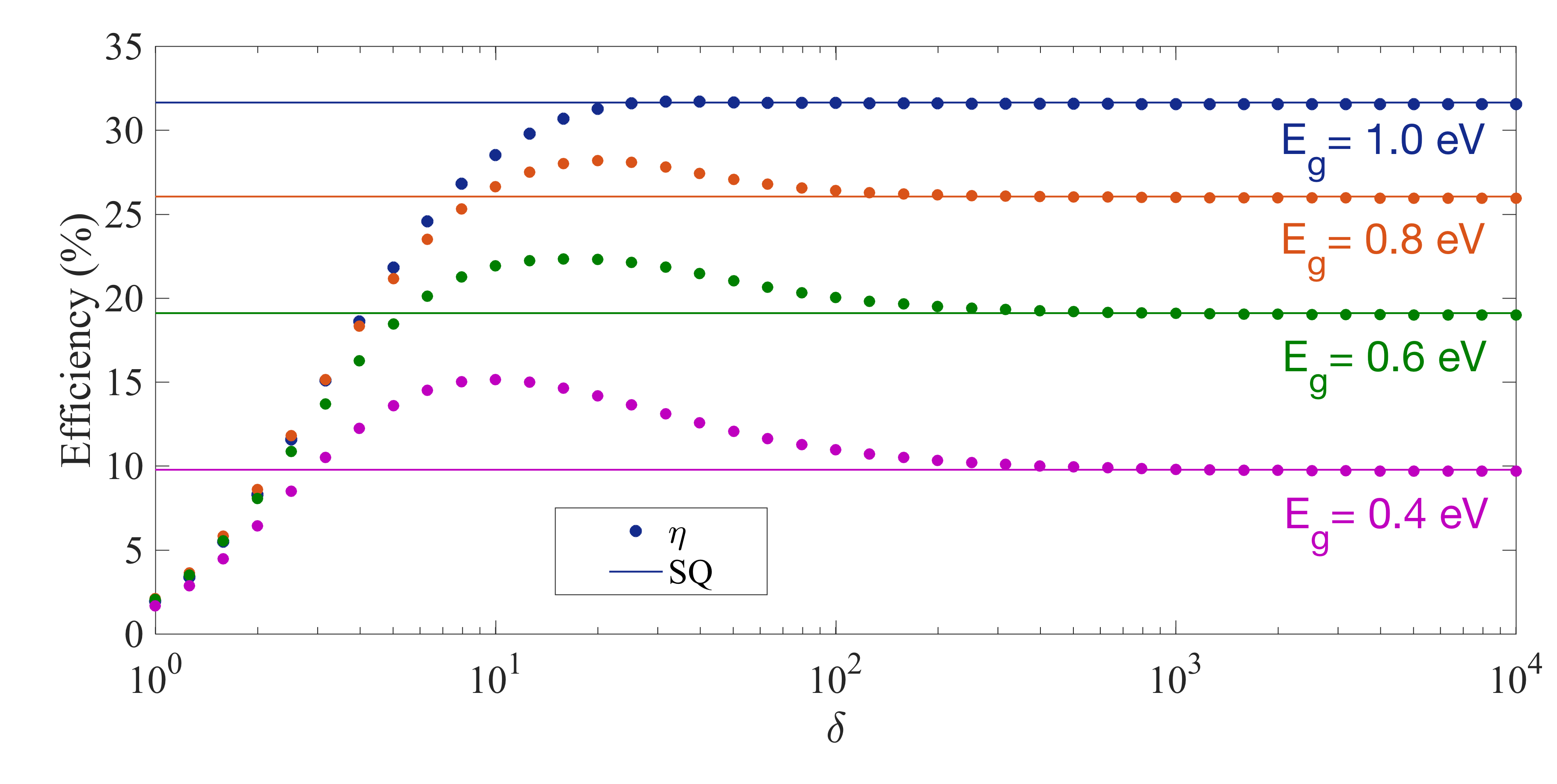}
\caption{\label{fig:deltadep}Calculated efficiencies for a range of $\delta$ values and a selection of band gaps, compared with the corresponding Shockley-Queisser limit.}
\end{figure}

To study the influence of the band gap on the likelihood of the efficiency exceeding the Shockley-Queisser limit, we calculate the efficiency for $\delta \in [1, 10^4]$ and over the band gap range $E_g~\in~[0.3, 3]~\si{\electronvolt}$. We show the $\delta$-dependency of the efficiency for a selection of band gap values in Fig.~\ref{fig:deltadep}. We can see that for low band gaps, the calculated efficiency crosses the detailed balance limit of the corresponding band gap, in order to return to the limit value for $\delta \rightarrow \infty$. Since $\delta$ can be related to the thickness of the material, this implies that for lower band gap materials, there is a thickness that is optimal for the efficiency. Moreover, a clear trend is visible, with the efficiency exceeding the Shockley-Queisser limit more as the band gap is decreased. This is also what we observe when we look at the plot for the maximum efficiency values in Fig.~\ref{fig:SLME}.

It is interesting to note that the SLME values of the materials that exceed the Shockley-Queisser limit are still below the maximum efficiency for the model absorptivity functions of the corresponding band gap in Fig.~\ref{fig:SLME}. However, this does not imply that the logistic function maxima curve represents a new upper limit. It is entirely possible that there is another function profile that would allow for higher efficiencies. Using the logistic function approach, we are simply able to observe for which band gap range the Shockley-Queisser limit does not provide a theoretical upper limit.

\section{Conclusion}

In their 1961 paper, Shockley and Queisser characterized their calculated efficiency as an upper limit, because of the assumption that if every photon with an energy above the band gap is absorbed, the obtained efficiency must be maximal. Although this assumption may seem entirely sensible at first glance, it does not consider the fact that it also maximizes the recombination current, which is calculated using the detailed balance principle. Because an increased recombination results in a lower efficiency, this means that lowering the absorptivity can produce higher efficiencies than the Shockley-Queisser limit under the right conditions. By using a model absorptivity function, which closely resembles absorptivity spectra calculated from first principles, we have shown that this can occur for low band gaps. This means that one must take care when dismissing low band gap materials based on their Shockley-Queisser limit, for their actual efficiency at certain thicknesses might still make them suitable for thin film photovoltaic applications.

\bibliography{sqlimit}

\begin{thebibliography}{14}
\expandafter\ifx\csname natexlab\endcsname\relax\def\natexlab#1{#1}\fi
\expandafter\ifx\csname url\endcsname\relax
  \def\url#1{\texttt{#1}}\fi
\expandafter\ifx\csname urlprefix\endcsname\relax\def\urlprefix{URL }\fi

\bibitem[{Shockley \& Queisser(1961)}]{Shockley1961DetailedCells}
Shockley, W. \& Queisser, H.~J.
\newblock {Detailed Balance Limit of Efficiency of p‐n Junction Solar Cells}.
\newblock \emph{J. Appl. Phys.} \textbf{32}, 510--519 (1961).

\bibitem[{Yu \& Zunger(2012)}]{Yu2012IdentificationMaterials}
Yu, L. \& Zunger, A.
\newblock {Identification of Potential Photovoltaic Absorbers Based on
  First-Principles Spectroscopic Screening of Materials}.
\newblock \emph{Phys. Rev. Lett.} \textbf{108}, 068701 (2012).

\bibitem[{Meng \emph{et~al.}(2016)}]{Meng2016AlloyingApplication}
Meng, W. \emph{et~al.}
\newblock {Alloying and Defect Control within Chalcogenide Perovskites for
  Optimized Photovoltaic Application}.
\newblock \emph{Chem. Mater.} \textbf{28}, 821--829 (2016).

\bibitem[{Lee \emph{et~al.}(2014)Lee, Lee, Oh, Kim \& {Chang
  KJ}}]{Lee2014ComputationalCrystals}
Lee, I.-H., Lee, J., Oh, J.~O., Kim, S. \& {Chang KJ}.
\newblock {Computational search for direct band gap silicon crystals}.
\newblock \emph{Phys. Rev. B} \textbf{90}, 115209 (2014).

\bibitem[{Bercx \emph{et~al.}(2016)Bercx, Sarmadian, Saniz, Partoens \&
  Lamoen}]{Bercx2016First-principlesSilicon}
Bercx, M., Sarmadian, N., Saniz, R., Partoens, B. \& Lamoen, D.
\newblock {First-principles analysis of the spectroscopic limited maximum
  efficiency of photovoltaic absorber layers for CuAu-like chalcogenides and
  silicon}.
\newblock \emph{Phys. Chem. Chem. Phys.} \textbf{18}, 20542--20549 (2016).

\bibitem[{Sarmadian \emph{et~al.}(2016)Sarmadian, Saniz, Partoens \&
  Lamoen}]{Sarmadian2016First-principlesChalcogenides}
Sarmadian, N., Saniz, R., Partoens, B. \& Lamoen, D.
\newblock {First-principles study of the optoelectronic properties and
  photovoltaic absorber layer efficiency of Cu-based chalcogenides}.
\newblock \emph{J. Appl. Phys.} \textbf{120}, 085707 (2016).

\bibitem[{Nelson \emph{et~al.}(2013)}]{Nelson2013ExceedingConversion}
Nelson, C.~A. \emph{et~al.}
\newblock {Exceeding the Shockley–Queisser limit in solar energy conversion}.
\newblock \emph{Energy Environ. Sci.} \textbf{6}, 3508 (2013).

\bibitem[{Shah \emph{et~al.}(2004)}]{Shah2004Thin-filmTechnology}
Shah, A.~V. \emph{et~al.}
\newblock {Thin-film silicon solar cell technology}.
\newblock \emph{Progress in Photovoltaics: Research and Applications}
  \textbf{12}, 113--142 (2004).

\bibitem[{Heremans \emph{et~al.}(2009)Heremans, Cheyns \&
  Rand}]{Heremans2009StrategiesArchitecture}
Heremans, P., Cheyns, D. \& Rand, B.~P.
\newblock {Strategies for Increasing the Efficiency of Heterojunction Organic
  Solar Cells: Material Selection and Device Architecture}.
\newblock \emph{Acc. Chem. Res.} \textbf{42}, 1740--1747 (2009).

\bibitem[{K{\"{o}}nig \emph{et~al.}(2010)}]{Konig2010HotDesign}
K{\"{o}}nig, D. \emph{et~al.}
\newblock {Hot carrier solar cells: Principles, materials and design}.
\newblock \emph{Physica E: Low-dimensional Systems and Nanostructures}
  \textbf{42}, 2862--2866 (2010).

\bibitem[{Hanna \& Nozik(2006)}]{Hanna2006SolarAbsorbers}
Hanna, M.~C. \& Nozik, A.~J.
\newblock {Solar conversion efficiency of photovoltaic and photoelectrolysis
  cells with carrier multiplication absorbers}.
\newblock \emph{J. Appl. Phys.} \textbf{100}, 074510 (2006).

\bibitem[{201(2012)}]{2012ASTMPA}
{ASTM G173-03(2012), Standard Tables for Reference Solar Spectral Irradiances:
  Direct Normal and Hemispherical on 37{${}^\circ$} Tilted Surface, ASTM
  International, West Conshohocken, PA} (2012).
\newblock \urlprefix\url{www.astm.org}.

\bibitem[{Cuevas(2014)}]{Cuevas2014TheJ0}
Cuevas, A.
\newblock {The Recombination Parameter J0}.
\newblock \emph{Energy Procedia} \textbf{55}, 53--62 (2014).

\bibitem[{R{\"{u}}hle(2016)}]{Ruhle2016TabulatedCells}
R{\"{u}}hle, S.
\newblock {Tabulated values of the Shockley–Queisser limit for single
  junction solar cells}.
\newblock \emph{Sol. Energy} \textbf{130}, 139--147 (2016).

\end{thebibliography}
\bibliographystyle{nature}

\end{document}